\documentclass[10pt]{article}

\usepackage{cite} 

\usepackage{hyperref}
\usepackage{graphicx}
\usepackage[font=scriptsize]{caption}

\usepackage[a4paper,top=2cm,bottom=2cm,left=3cm,right=3cm]{geometry} 

\begin{document}

\title{\textbf{The usage of Kahoot! during activities for schools on physics}\vspace{-4ex}}
\date{}
\maketitle
\begin{center}

\author{Massimiliano Di Blasi$^{1}$, Ilaria De Angelis$^{1,2}$, Adriana Postiglione$^{1,2}$}
\end{center}

\paragraph{} \parbox[t]{1\columnwidth}{$^1$Dipartimento di Matematica e Fisica, Universit\`a degli Studi Roma \\Tre, Rome (Italy)\\%
    $^2$INFN Sezione di Roma Tre, Rome, (Italy)\\}

\begin{abstract}
The transition of many educational activities from in-presence to
online, due to the Covid-19 emergency, has caused various difficulties to students,
especially with regards to maintaining concentration. Even and especially in online
mode, it is thus important to pay attention to the interaction with participants. In
this context, the Department of Mathematics and Physics of Roma Tre University
has proposed a series of activities dedicated to schools that have proven to be able
to guarantee the engagement of the participants, also thanks to the game-based
learning platform \textit{Kahoot!}.

\end{abstract}


\section{Introduction}
The Covid-19 emergency led to the development of remote teaching proposals. This
caused multiple obstacles to the learning process, especially in terms of motivation, stress
and academic progress of students \cite{Klein}. Emblematic is the fact that it is not uncommon
to see students logging on and, after a while, walking away from the computer. For
these reasons, it is important to find a way to engage all participants as much as possible
and to actively involve students \cite{Freeman, Prince, Hake} avoiding teaching activities structured as mere
frontal lessons. This not only applies to educational activities carried out in schools or
universities, but also to all initiatives that aim to bring students closer to the world of
science, such as, for example, the activities that universities offer to high school students.
It is in this context that at the Department of Mathematics and Physics of the Rome
Tre University we have developed activities for high school students that have ensured
a strong interaction with the participants. In particular, we used different approaches
such as discussions, questioning, inquiry-based exercises, interactive use of technology,
software and also playful quizzes that made use of \textit{Kahoot!}\footnote{Kahoot! is a game-based learning platform: \url{kahoot.com}}, a tool that allowed us to
keep students’ concentration high also thanks to its competitive factor \cite{Wang}. In this paper, we will describe the online activities for high school students proposed
by the Department, the ways of engagement we have found and the feedback we have
received.

\section{Our online interactive proposal}
The online activities for students proposed by the Department involved 8 groups of
students, including 7 classes and a mixed group of students from different schools, for
a total of 170 high school students, and lasted from September 2020 to May 2021. A
dedicated course was created for each group: the aim was to bring participants closer to
physics and mathematics through topics they do not deal with at school or to deepen
those topics that already known from a university point of view.
Each course lasted from 20 to 30 hours and was organized in two-hour afternoon
meetings through the Zoom platform, which school teachers were also invited to attend
to as auditors. To facilitate students’ intervention during the meetings, we opted for the
meeting mode and avoided the webinar one, to allow them to turn on the microphone
and the webcam at any time. Then, during the meetings, students were often questioned
about their impressions on a given activity. Furthermore, to ensure active involvement
of introverted students as well, we also proposed some simple exercises through special
software or platforms to discover knowledge through direct experience.
Another way we used to keep students’ attention high and get them involved was
playing with them using \textit{Kahoot!}, a point-based quiz, where students have limited time
to answer and, after each question, can see a partial leaderboard \cite{Wang}. During the game,
students can use nicknames, which empowers the interaction of all students, even if not
particularly outgoing; moreover, scoring depends not only on correct answers, but also
on how fast they are given.
Meetings with each group, albeit on different topics, all followed the same structure:
after a first part dedicated to greetings, questions and curiosities, we introduced the topic
that we would have dealt with that day. Then the lesson was structured through theoretical and practical activities carried out through web platforms or recorded videos\footnote{ The YouTube playlist created by the Department can be found at \url{https://www.youtube.com/channel/UCl3bqukQQ-VEpJBGERfbEoQ/playlists?view=50&sort=dd&shelf id=4.}}.
All the activities were designed to be divided into blocks of about 15 minutes; in order
to intersperse these blocks, we thus opened a short discussion on the topics just treated
starting from the students’ questions and then providing a deeper insight with specific
reflections. At the end of each block, we used \textit{Kahoot!}. Specifically, students had to
answer three or four multiple-choice questions, by clicking the selected option directly
from their mobile phone. Usually, we used three blocks and eleven \textit{Kahoot!} questions
for each meeting. The questions we asked through \textit{Kahoot!} were asked in such a way as to require
an immediate answer from students (sometimes the right answer exactly reported the
same words or phrases used by the teacher). In fact, we chose to set 30 seconds as
the maximum time to answer the question; this also ensured greater competition among
participants.
After each question, we briefly showed and discussed the correct answer and commented on the wrong one. In an informal and playful atmosphere, partial ranking was
shown, and those who answered well and faster were praised, while the others were encouraged to do better and climb the ranking. In this way, we supported competition
between participants in a friendly way in \textit{Kahoot!} style. At the end of the match, the
final leaderboard was shown and the winner was acclaimed, and a new challenge opened
up to all peers for the next meeting.
In the final meeting of the course, an anonymous evaluation questionnaire about the
entire proposal was presented to participants.

\section{Feedback we received}
147 students out of 170 students who participated in our courses answered the anonymous evaluation questionnaire proposed at the end. Overall, the feedback we received
was very positive.
In fact, as regards the questions relating to satisfaction about the proposed course,
almost all participants (91.16
think it was worth participating?” 49 answered “absolutely yes”, 52 “yes”, 33 “yes rather
than no”, 10 “no”, 3 “absolutely not”). The same percentage of students (91.16
claimed they would recommend it to their peers (“Would you recommend the course to
a peer of your age?” 96 answered “yes”, 22 “absolutely yes”, 16 “yes rather than no”, 5
“maybe”, 7 “no” and 1 student did not answer).
The active participation of the students during the course is also demonstrated by
the answers given to the \textit{Kahoot!} quizzes proposed during the lessons, which show that
most of the participants stayed active during the whole meeting. For example, taking
into consideration the analysis of a lesson common to all courses, it can be seen that
89.7\% of the students (139 out of 155 participants) answered all the questions asked, and
the 70.7\% of their answers was correct.
In addition to these results, we also had positive feedback from the teachers of the
classes involved. They told us not only that their students had a lot of fun and enjoyed
the whole course, but also that themselves were satisfied, so much that they would have
used both the materials we have proposed during our course and \textit{Kahoot!} with their
pupils.

\section{Conclusions}
Distance lessons have become an important part of school activities due to the Covid19 emergency, but this modality puts a strain on maintaining attention from students.
In this paper, we described a series of online activities aimed at 8 different groups of
students dealing with topics of physics and mathematics, in which participants showed
to remain effectively active and engaged also thanks to the use of \textit{Kahoot!}.
We indeed have seen that more than 91\% of the participants not only appreciated our activity but also would recommend it to a peer. Moreover, we received positive feedback
also from their teachers.
In particular, in our case, the use of \textit{Kahoot!} has been very useful for many reasons.
It helped students to maintain concentration, since they knew that there would have
been a quiz after each part of the lesson. Then, the existence of a final ranking increased
the attention of the participants pushing them to note even the most detailed insights.
Furthermore, \textit{Kahoot!} helped to enhance involvement and interaction among students,
as anonymity made it easier also for the shyest to get involved and allowed us to repeat
and highlight the main concepts covered during our lessons. Moreover, \textit{Kahoot!} also gave us the possibility of receiving feedback about students behaviour in real-time and checking if students paid attention for the full duration of the lesson. Specifically, it
allowed us to keep track of the degree of participation: for example, from the analysis
of a typical activity (the lesson common to all courses), we observed that about 90
participants answered all the questions asked with about 70
the short time allowed to respond. These results are particularly important if taking into
account that online mode could lead students to log in and then not really follow the
lesson.
Furthermore, since our data also give us a clue as to the understanding of the treated
topics in the short term, we think that the usage of \textit{Kahoot!} could be considered as a
valuable tool to study students’ comprehension also during ordinary lessons in presence.
It could be interesting to deepen this aspect in the future also thanks to the feedback we
could receive from school teachers who will adopt \textit{Kahoot!} in their lessons.

\section*{Acknowledgements}
This work was supported by the Italian Project Piano "Lauree Scientifiche". The
authors also want to thank Simonetta Pieroni, film-maker, for the excellent production
of the videos used during the described activity.

\end{document}